\documentclass[prl,twocolumn,showpacs,tightenlines,
preprintnumbers,amsmath,amssymb]{revtex4}
\usepackage{graphicx}
\usepackage{epstopdf}
\newcommand{\eq}[1]{(\ref{#1})}
\newcommand{\etal} {{{\it et al.\/}}}
\begin{document}
\title{Nonlinear phenomenology from quantum mechanics: Soliton in a lattice}
\author{Juha Javanainen}
\author{Uttam Shrestha}
\affiliation{Department of Physics, University of Connecticut,
Storrs, CT 06269-3046}
\date{\today}
\begin{abstract}
We study a soliton in an optical lattice holding bosonic atoms quantum mechanically using both an exact numerical solution and Quantum Monte Carlo (QMC) simulations. The computation of the state is combined with an explicit account of the measurements of the numbers of the atoms at the lattice sites. In particular, importance sampling in the QMC method arguably produces faithful simulations of the outcomes of individual experiments. Even though the quantum state is invariant under lattice translations, an experiment may show a noisy version of the localized classical soliton.
\end{abstract}
\pacs{03.75.Lm,03.65.Ta,11.30.Qc}
\maketitle
\raggedbottom
Quantum mechanics is widely accepted as the fundamental framework for physics. Nonetheless, a variety of nonlinear theories have met with unquestionable success in the modeling of macroscopic (many-body) systems. This state of the affairs begs for the question of how linear quantum mechanics mimics nonlinear behavior. Consider as an example a soliton of light propagating in a nonlinear optical fiber. Full quantum theory has solutions that in many respects resemble classical solitons~\cite{LAI89}, but finding such a quantum soliton does not solve the problem of linear versus nonlinear dynamic. For instance, in a translationally invariant optical fiber the quantum ground state may be chosen to be translationally invariant. Why is it, then, that localized solitons are so strongly favored by Nature that they form seemingly spontaneously?

It has been suggested that measurements play a part in the answer~\cite{KAN05}, but our tenet goes much further: Operationally, nonlinear phenomenology is created by the process of measurement. In this Letter we demonstrate our views by discussing bosonic atoms in an optical lattice, a problem that commands much current interest in its own right~\cite{MOR06,LEW07} and also presents a discrete-space version of the soliton in nonlinear optics. We solve the quantum mechanical ground state either by exact diagonalization or by Quantum Monte Carlo (QMC) simulations~\cite{HIR82,HEB05}, but with a measurement theoretical twist. Namely, while a run of a QMC simulation produces the numbers of the atoms at each lattice site to be used in calculations of various ground-state expectation values, we argue that these occupation numbers are also a faithful simulation of what one would find in a single experiment that measures the occupation numbers. When classically one expects a soliton, each run of the QMC simulation, and hence each individual experiment, also shows a distribution of the atoms over the lattice sites like a classical soliton.

Quantum mechanically, we have the Bose-Hubbard model with the Hamiltonian
\begin{equation}
\hat{H} = \hbar\sum_k \left[-\frac{\delta}{2}(\hat b^\dagger_{k+1}\hat b_k+\hat b^\dagger_{k-1}\hat b_k) +\frac{\kappa}{2}\, \hat b^\dagger_k\hat b^\dagger_k\hat b_k\hat b_k\right].
\label{QMH}
\end{equation}
Here $\hat b_k$ annihilates an atom at the site $k=0,\ldots,L-1$, $\delta$ is the amplitude for site-to-nearest-site tunneling, and $\kappa$ characterizes on-site atom-atom interactions. Our $L$-site lattice is periodic so that $k=L$ and $k=0$ are the same site.  This boundary condition, which would physically correspond to a ring lattice, is specifically chosen to make the Hamiltonian invariant under lattice translations. The total number of atoms $\hat N = \sum_k \hat b^\dagger_k \hat b_k$ is a constant of the motion, and its value is denoted by $N$.

The corresponding classical Hamiltonian is
\begin{equation}
H = \hbar\sum_k \left[-\frac{\delta}{2}(b^*_{k+1} b_k+\ b^*_{k-1} b_k) +\frac{\kappa}{2}\, |b_k|^4\right].
\label{CLH}
\end{equation}
Here $b_k$ are complex numbers such that $|b_k|^2$ stands for the number of atoms at the site $k$. The quantities $b_k$ and $b^*_{k'}$ are regarded as classical canonical conjugates with the Poisson brackets $\{b_k,b^*_{k'} \}=-(i/\hbar)\delta_{k,k'}$. Hamilton's equation of motion for $b_k$, 
\begin{equation}
i\dot b_k = -\frac{\delta}{2}(b_{k+1}+b_{k-1}) + \kappa |b_k|^2 b_k\,,
\label{DNLSE}
\end{equation}
the lattice analog of the Gross-Pitaevskii equation,  is commonly called Discrete Nonlinear Schr\"odinger equation (DNLSE). The total atom number $N=\sum_k |b_k|^2$ is again a constant of the motion.

The ground state of the classical Hamiltonian~\eq{CLH} is found, e.g., by integrating the DNLSE~\eq{DNLSE} in imaginary time. For repulsive interactions between the atoms, $\kappa\ge0$, the lowest-energy state is spatially uniform, $b_k=\sqrt{N}e^{i\varphi}$, where $\varphi$ is a global phase. On the other hand, for a sufficiently strong attractive interaction $\kappa<0$, the lowest-energy state is a soliton localized around some lattice site~\cite{SCO83}. A soliton is a stationary state of the DNLSE; the amplitudes evolve in time according to $b_k(t)=e^{-i\mu t/\hbar}b_k(0)$, where $\mu$ is the chemical potential. Physically, the soliton reflects a balance between the tendencies of the atoms to collect together drawn by the attractive interactions, and to disperse as a result of the site-to-site hopping. A lattice-translated solitonic ground state is also a ground state, but a nontrivial superposition of ground-state solitons is not. Nonlinearity makes localization an inescapable feature of the solitonic ground state.

In search of the quantum counterpart of a soliton we study the Hamiltonian~\eq{QMH}. Introduce the thermal density operator at the inverse temperature $\beta$, $\hat\rho(\beta) = e^{-\beta\hat H}/{\rm Tr}[e^{-\beta\hat H}]$, the number operators of the sites $k$, $\hat n_k=\hat b^\dagger_k\hat b_k$, the collection of these operators $\hat{\bf n}=\{\hat n_0,\ldots,\hat n_{L-1}\}$, and the simultaneous eigenstates $|{\bf n}\rangle$ of the number operators $\hat{\bf n}$  characterized by a collection of $L$ integers ${\bf n}=\{n_0,\ldots,n_{L-1}\}$. We call such a set ${\bf n}$ of the occupations numbers $n_k$  a number state of the lattice. As the operators $\hat{n}_k$ commute, according to standard quantum mechanics their values can be measured simultaneously. We say that the operator $\hat{\bf n}$ is measured. Any function $f({\bf n})$ of the number state defines in a natural way the operator function ${f}(\hat{\bf n})$, and the thermal expectation value of ${f}(\hat{\bf n})$ equals
\begin{equation}
\langle{f}(\hat{\bf n})\rangle = \sum_{\bf n}P({\bf n}) f({\bf n});\quad P({\bf n}) = \frac{\langle{\bf n}|e^{-\beta \hat H}|{\bf n}\rangle}{\sum_{\bf n}\langle{\bf n}|e^{-\beta \hat H}|{\bf n}\rangle}\,.
\label{EXPT}
\end{equation}
On the other hand, suppose that by some means, call it importance sampling, it is possible to produce number states ${\bf n}_m$, $m=1,2,\ldots$ at random, but with frequencies proportional to the corresponding $P({\bf n}_m)$ as in Eq.~\eq{EXPT}. The expectation value in Eq.~\eq{EXPT} may then also be written
\begin{equation}
\langle{f}(\hat{\bf n})\rangle=\langle f ({\bf n})\rangle = \lim_{M\rightarrow\infty}\frac{1}{M}\sum_{m=1}^{M} f({\bf n}_m)\,.
\label{CEXPT}
\end{equation}

Considering any quantity that can be expressed in terms of expectation values of functions of the form $f({\bf n})$, the values from quantum measurements and from importance sampling agree. This applies to average atom number at a site, correlations of atom numbers between the sites,  and so on: The measured values of the number operators $\hat n_k$  agree in every possible statistical characterization with the occupation numbers $n_k$ in the importance-sampled number states ${\bf n}_m$. One experiment and one instance of sampling each produce a random set of occupation numbers, but in repeated runs the statistics of these random numbers could be verified to be the same for the experiments and for importance sampling. We therefore formulate our key {\em interpretative hypothesis}: Each number state ${\bf n}_m$ coming from importance sampling is a representative outcome of an experiment measuring the operator $\hat{\bf n}$.

For brevity we focus on the low-temperature limit, $\beta\rightarrow\infty$. If the ground state of the Hamiltonian~\eq{QMH} is nondegenerate, the density operator is simply the projection onto the ground state. On the other hand, if the ground state is degenerate, the zero-temperature state is an equal mixture of the degenerate ground states. As a lattice translation is a symmetry operation of the Hamiltonian, it is possible to select all energy eigenstates so that they are invariant under  lattice translations. Hence, the unique zero-temperature density operator must be invariant under lattice translations. Unlike the classical soliton, the quantum mechanical ground state (or, for that matter, any thermal-equilibrium state) will not single out any particular lattice site.

We first take up the two-site lattice, $L=2$. Simple as the problem is, for parameters such as $N\kappa/\delta = -2.309$ we already have the conundrum that the classical ground state is a soliton with $\hbox{$\frac{3}{4}$}$ of the atoms in one site and $\hbox{$\frac{1}{4}$}$ in the other, whereas the quantum ground state is translation invariant. So, what would the experiments see?

 Here the quantum mechanical state space is spanned by the $N+1$ vectors $|n,N-n\rangle$, where $n$ ($N-n$) is the number of the atoms in the site $k=0$ ($k=1$). The two-site problem is trivial to solve numerically. For a fixed value of the parameter $N\kappa/\delta$ that classically gives a soliton, and in the limit $N\rightarrow\infty$, the ground state is doubly degenerate. Therefore we write the low-temperature ground state as a 50/50 statistical mixture of the two lowest-energy quantum states. As far as observed atom numbers are concerned, the complete statistics is determined by the probabilities $P_n$ that $n$ of the atoms are found in the $k=0$ site. 
 
\begin{figure}
 \begin{center}
\includegraphics[width=8.5cm]{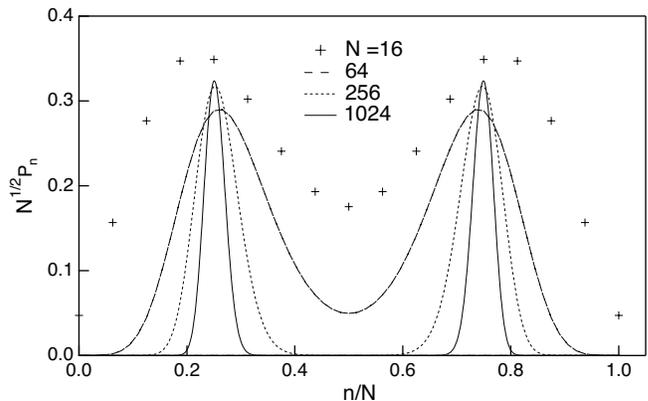}
\end{center}
\vspace{-20pt}
\caption{Representation of the probabilities $P_n$ that $n$ atoms are in one given site of a two-site lattice for different numbers of the atoms $N$, as indicated in the legend. This figure is from exact quantum solutions for the fixed value of the parameter $N\kappa/\delta=-2.309$. }
\label{FIG_1}    
\vspace{-10pt}
\end{figure}

 In Fig.~\ref{FIG_1} we plot the quantity $\sqrt{N}\,P_n$ as a function of the fraction of the atoms in the $k=0$ site, $n/N$. Here $N\kappa/\delta = -2.309$ is fixed, and the atom number $N$ is varied as indicated in the legend. Importance sampling would produce number states ${\bf n}=\{n,N-n\}$  at frequencies proportional to $P_n$. As Fig.~\ref{FIG_1} shows, with an increasing atom number $N$ the sampling, and the experiments, would give an increasingly accurate split of the atoms so that $\hbox{$\frac{3}{4}$}$ of the them is in one site and $\hbox{$\frac{1}{4}$}$  in the other, just like in the classical soliton. The reason for the $\sqrt{N}$ multiplier for the probabilities is that one may then easily see from Fig.~\ref{FIG_1} that the fluctuations in the $3:1$ split scales with the atom number $N$ like $1/\sqrt{N}$.
 
While the Hamiltonian is invariant under a lattice translation, an individual measurement result of the atoms numbers with the approximate $3:1$ split between the sites is not. However, either one of the two sites is the one with the higher occupation number with the same probability, so that averaging over the experiments restores the symmetry. 

The resemblance of this situation to spontaneously broken translation symmetry has been brought up in numerous papers on the continuous soliton problem and the attendant connection to measurement theory has also been noted~\cite{KAN05}, but here we take a very direct approach.

 Consider interference between two Bose-Einstein condensates, an analogous problem from the past~\cite{JAV96,CIR96,CAS97} that continues to evoke new angles~\cite{MUL06}. The traditional view is that gauge symmetry of the condensate is broken, which endows the BEC with a phase. The difference in the phases is observable when two condensates are made to overlap, which indeed produces an interference pattern~\cite{AND97}. 
In contrast, we have predicted an interference pattern for two number-state condensates without ever assuming any broken symmetry or phase~\cite{JAV96}. The key was to simulate the measurements of the positions of the atoms.
There are correlations between atomic positions in the state vector of the two overlapping condensates. Every time the position of one atom is observed, the state gets reduced so that it is compatible with the newly gained measurement result. This reduction brings out the correlations by modifying the position distribution for the next atom to be observed. Continuing in this way, the measured positions of the atoms produce an interference pattern. It comes about as a combination of two elements: correlations between atomic positions in the state, and measurements that convert the correlations into the observations.

Easy as the corresponding measurement simulation would be in the case of a two-site lattice~\cite{CAS97,JAV00}, we have not carried it out because the analog is clear enough as it is. Even if the ground state is translationally invariant, the potential for a soliton is there in the correlations between the positions of the atoms: The attractive interactions favor atoms collecting in the same site. However, it is the observations  that ultimately make the soliton.

With increasing numbers of atoms and lattice sites an exact numerical solution~\cite{NS} of the lattice system becomes impractical. Besides, the solution does not tell what an experiment would see; one has to add a measurement simulation along the lines of Refs.~\cite{JAV96,CAS97,JAV00}. QMC simulations may provide a sweeping answer to the questions of both finding the state and measuring it. This is because importance sampling is a core idea in QMC methods, and in Monte-Carlo simulations in general~\cite{BIN02}.

Our QMC simulations take place on a $L\times N_\beta$ grid~\cite{FOOT1}, where the first dimension represents the lattice sites and the second dimension corresponds to $N_\beta$ steps in inverse temperature with the size $\beta/N_\beta$.  We use an elementary world line algorithm combined with the checkerboard decomposition of the Hamiltonian~\cite{HIR82, HEB05}. The grid is initially seeded with the same number state of the lattice at each step of the inverse temperature.  A site on the grid is selected in a suitable way, say, at random, and at the site a move is proposed, also at random;  an atom moves left or right to or from an adjacent lattice site. The move is either accepted or rejected with a probability that is calculated from the state of the grid in the immediate neighborhood of the chosen site.
Moves are attempted in this way until the grid has relaxed to a steady state and executes fluctuations around it. We use periodic boundary conditions in the inverse-temperature direction as well, so that a number state ${\bf n}$  in the relaxed grid at any step of the inverse temperature qualifies as a sample drawn at a frequency proportional to $P({\bf n})$, Eq.~\eq{EXPT}.  Overall, a QMC simulation sets up an artificial dynamics that has nothing to do with the real dynamics of the lattice, but realizes importance sampling for the thermal state.

To demonstrate spontaneous breaking of translation symmetry we would like to initialize the grid so that every site has the same atom number. This works for small numbers of atoms and lattice sites, and for suitable strengths  of the attractive interaction solitons are found once the simulation has converged. However, a soliton is nucleated from fluctuations of atom numbers in the grid, and for large atom numbers a sufficiently large fluctuation does not occur during the number of attempted moves we are willing to wait out. In such a case we form the soliton by seeding the grid in the direction of the lattice sites with a narrow distribution of the atoms.

\begin{figure}
 \begin{center}
\includegraphics[width=8.5cm]{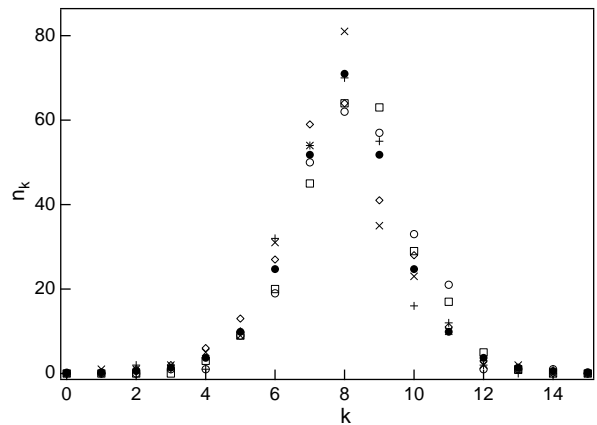}
\end{center}
\vspace{-20pt}
\caption{Atom numbers $n_k$ at lattice sites $k$ for the classical soliton (solid circles) and for five solitons sampled from quantum mechanics; the symbols for the six solitons are all resolved at the site $k=9$. The parameters are $L=16$, $N=256$, and $\kappa/\delta=-0.004$. The temperature is $\delta/100$ in units of frequency. Given the occupation numbers for the soliton $n_k^C$ and for the QMC results $n_k^Q$, the $\ell^2$ distances between the soliton and the five simulations, $d=[\sum_k(n_k^Q-n_k^C)^2]^{1/2}$, equal 12.8, 20.9, 18.8, 17.8, and 15.8.}
\label{FIG_2}    
\vspace{-10pt}
\end{figure}

Let us now compare the classical soliton and results from quantum simulations  for  $L=16$ lattice sites and $N=256$ atoms. We choose the interaction strength $\kappa/\delta=-0.004$, for which there is a classical soliton that spans several sites but is well contained inside the lattice length $L$. In Fig.~\ref{FIG_2} we plot both the classical soliton (solid circles) and five importance-sampled solitons from the QMC method (various symbols, best resolved at the lattice site $k=9$). QMC results have been translated along the lattice so that the root-mean square difference from the classical soliton is minimized. The fluctuations in the simulations are rather large, but an eye would recognize every result of sampling as a soliton.

We have discussed explicitly two examples summarized in Figs.~\ref{FIG_1} and~\ref{FIG_2}, but in our experience the results are generic:  If for some fixed values of $L$ and $N\kappa/\delta$ there is a classical soliton, in the limit of large $N$ a measurement on the quantum system will produce a soliton as well.

Our main technical point that each outcome from importance sampling faithfully represents an outcome of a single experiment may seem self-evident, but  in practice it appears to be difficult to grasp this type of an argument. We therefore expand on the underlying philosophy.

Ultimately, the claim is that the statistics of a random variable, here ${\bf n}$, determines how the realizations look like. We have resorted to the same idea before~\cite{JAV86,JAV91}, for instance when we predicted ``quantum jumps'' in the intensity of  the light scattered off a three-level system with a long-lived ``shelving state'' ab initio from quantum mechanics~\cite{JAV86}. The intensity correlation functions for the scattered light are continuous functions of their time arguments, but they are the correlation functions of a Markov jump process. Hence, the light intensity should look like a Markov jump process, i.e., execute abrupt jumps. These quantum jumps, originally envisaged by Dehmelt~\cite{DEH75}, were subsequently observed~\cite{DEH86} and are now a standard laboratory technique for detecting weak transitions. In the present case we have no prior experience with the underlying random variable, but importance sampling shows that the realizations look like solitons. Hence, experiments should produce solitons.

To further illustrate the paradigm shift in our reasoning we compare with traditional QMC simulations. Ordinarily one would produce a number of samples and use them to calculate average quantities such as occupation numbers of the sites, standard deviations of the occupation numbers, correlations in atom numbers between adjacent sites, and so on; and maybe compare such averages with averages extracted from experiments. However, such a process may miss the forest from the trees: Even though the results from importance sampling display solitons, solitons may be difficult to uncover from the averaged quantities.

Recognizing the soliton as the basic outcome from measurements should inspire novel theoretical work. Consider analyzing fluctuations of a soliton using linearization of quantum mechanics around the classical soliton \cite{DZI04}, which in the context of Bose-Einstein condensates is commonly called Bogoliubov theory. For a fixed number of sites the DNLSE can be rewritten so that the only dimensionless parameter  containing the atom number is $N\kappa/\delta$, the same applies to the linearization of  the DNLSE that gives the Bogoliubov theory, and therefore also to atom number fluctuations around the classical soliton~\cite{JAV96b,GAR97}. We have here the interesting situation that the atom number fluctuations in the Bogoliubov theory can only depend on atom number through the parameter $N\kappa/\delta$, which directly contradicts our Fig.~\ref{FIG_1}.

There are many layers in our discussion. At the lowest level we introduce the measurement-theoretical idea that importance sampling in our simple QMC method  literally simulates the experiments. New ways of comparing theory and experiment then open up. We apply this line of thought  to predict localized solitons for atoms in an optical lattice from quantum mechanics in spite of the fact that the quantum state is translationally invariant and does not favor any particular lattice site. At the top level we promote the notion that  classical nonlinear phenomena in a macroscopic system are implicit in the correlations within the quantum state, but measurements are the agent that ultimately brings them forth.

This work is supported in part by NSF (PHY-0750668).


\vspace{-20pt}

\begin{references}
\bibitem{LAI89} Y. Lai and H. A. Haus, \pra {\bf 40} 844 (1989); \pra {\bf 40} 854 (1989).
\bibitem{KAN05} R. Kanamoto \etal,  \prl {\bf 94}, 090404 (2005); \pra {\bf 73}, 033611 (2006).
\bibitem{MOR06} O. Morsch and M. Oberthaler, \rmp {\bf 78}, 179 (2006).
\bibitem{LEW07} M. Lewenstein \etal, Advances in Physics {\bf 56}, 243 (2007).
\bibitem{HIR82} J.\ E.\ Hirsch \etal, Phys.\ Rev.\ B {\bf 26}, 5033 (1982).
\bibitem{HEB05} F. Hebert \etal, \pra {\bf 71}, 063609 (2005).
\bibitem{SCO83} A. C. Scott and L. Macneil, Phys. Lett. A {\bf 98}, 87 (1983).
\bibitem{JAV96}  J. Javanainen and  S. M. Yoo, \prl {\bf 76}, 161 (1996).
\bibitem{CIR96}  J. I. Cirac \etal,  \pra {\bf 54}, R3714 (1996).
\bibitem{CAS97} Y. Castin and J. Dalibard, \pra {\bf 55}, 4330 (1997).
\bibitem{MUL06} W. J. Mullin \etal, \pra {\bf 74}, 023610 (2006).
\bibitem{AND97}  M. R. Andrews \etal, Science {\bf 275}, 637 (1997).
\bibitem{JAV00} J. Javanainen, J. Phys. B {\bf 33},  5493 (2000).
\bibitem{NS} M. W. Jack and M. Yamashita, Phys. Rev. A {\bf 71}, 023610 (2005); N. Oelkers and J. Links, \prb {\bf 75}, 115119 (2007).
\bibitem{BIN02} K. Binder and D. W. Heermann, {\it Monte Carlo simulations in statistical physics: An introduction}, 4th Ed. (Springer, Berlin, 2002).
\bibitem{FOOT1} Even though conceptually the grid size is $L\times N_\beta$, for algorithmic reasons the actual size is $L\times2N_\beta$.
\bibitem{JAV86} J. Javanainen, \pra {\bf 33}, 2121 (1986).
\bibitem{JAV91} J. Javanainen, Phys. Lett. {\bf A161}, 207 (1991).
\bibitem{DEH75} H. Dehmelt, Bull. Am. Phys. Soc. {\bf 20}, 60 (1975).
\bibitem{DEH86} W. Nagourney \etal, \prl {\bf 56}, 2797 (1986); J. C. Bergquist \etal, \prl {\bf 57}, 1699 (1986).
\bibitem{DZI04} J. Dziarmaga, \pra {\bf 70}, 063616 (2004).
\bibitem{JAV96b} J. Javanainen, \pra {\bf  54}, R3722 (1996).
\bibitem{GAR97} C. W. Gardiner, \pra {\bf  56}, 1414 (1997).
\end{references}
\end{document}